\documentclass[aps,prb,twocolumn,preprintnumbers,amsmath,amssymb]{revtex4}
\usepackage{graphicx}
\usepackage{xcolor}
\begin{document}
\title{Physical properties of MnSi at extreme doping with Co: Quantum criticality.}

\author{A.~E.~Petrova}
\affiliation{P.~N.~Lebedev Physical Institute, Leninsky pr., 53, 119991 Moscow, Russia}

\author{S.~Yu.~Gavrilkin}
\affiliation{P.~N.~Lebedev Physical Institute, Leninsky pr., 53, 119991 Moscow, Russia}
\author{A~.Yu.~Tsvetkov}
\affiliation{P.~N.~Lebedev Physical Institute, Leninsky pr., 53, 119991 Moscow, Russia}
\author{Dirk Menzel}
\affiliation{Institut f\"{u}r Physik der Kondensierten Materie, Technische Universit\"{a}t Braunschweig, D-38106 Braunschweig, Germany}
\author{Julius Grefe}
\affiliation{Institut f\"{u}r Physik der Kondensierten Materie, Technische Universit\"{a}t Braunschweig, D-38106 Braunschweig, Germany}
\author{S.~Khasanov}
\affiliation{Institute for Solid State Physics of RAS, Chernogolovka, Russia}
\author{S.~M.~Stishov}
\email{stishovsm@lebedev.ru}
\affiliation{P. N. Lebedev Physical Institute, Leninsky pr., 53, 119991 Moscow, Russia}

\begin{abstract}
The samples of (Mn$_{1-x}$Co$_x$)Si with $x=0.15$ and $x=0.17$ were grown and their physical properties: magnetization and magnetic susceptibility, resistivity and heat capacity were studied. The data analysis included also the previous results at $x=0.057, 0.063, 0.09$. The indicated doping MnSi with Co completely destroys the helical phase transition whereas basically saves the helical fluctuation area normally situated slightly above the phase transition temperature. This area spreading from $\sim$5 to 0 K is not changed much with doping and forms some sort of helical fluctuation cloud revealing the quantum critical properties: $C_p/T\rightarrow\infty$ at $T\rightarrow0$.
\end{abstract}
\maketitle

\section{Introduction}
The study of quantum criticality in the model substance of MnSi by doping with Fe and Co were carried out in Ref.~\cite{1,2,3,4}. The main idea was to imitate the high-pressure condition by substituting Mn with smaller ions that would place the material in a situation where the pure MnSi experiences a quantum phase transition at zero temperature~\cite{5,6,7}.  As was shown in Ref.~\cite{1,2} the helical phase transition in (Mn$_{1-x}$Fe$_x$)Si can be seen close to zero temperature at doping level $x\approx 0.16 -0.19$. At the same time in case of (Mn$_{1-x}$Co$_x$)Si doping MnSi with Co completely destroys the phase transition at $x\approx 0.04$ and temperature about 8~K. However, the helical fluctuation maxima or humps, which are normally situated slightly above the phase transition temperature, survive even much heavier doping~\cite{4}. Although, temperature of the mentioned maxima appears to be not quite sensitive to doping above $x\approx 0.05$. This intriguing feature led to a tentative conclusion, which should be verified, that at large concentration of Co substitutes a cloud of the helical fluctuations spreading over a significant range of concentrations and temperatures arose close to 0~K (Ref.~\cite{4}).

\section{Experimental results and discussion}
With all that in mind we decided to extend our previous study~\cite{4} by including samples with larger concentration of Co dopant that probably could throw a new light to this problem.  The samples of (Mn$_{1-x}$Co$_x$)Si with $x>0.05$ were prepared by a procedure, described in Ref.~\cite{4} and two of them with $x=0.15$ and $x=0.17$, as determined by the electron probe microanalysis, were selected for further examination. The lattice parameters of the samples, measured by the X-ray powder diffraction techniques, are correspondingly 4.5464~\AA\ and 4.5457~\AA.

\begin{figure}[htb]
\includegraphics[width=80mm]{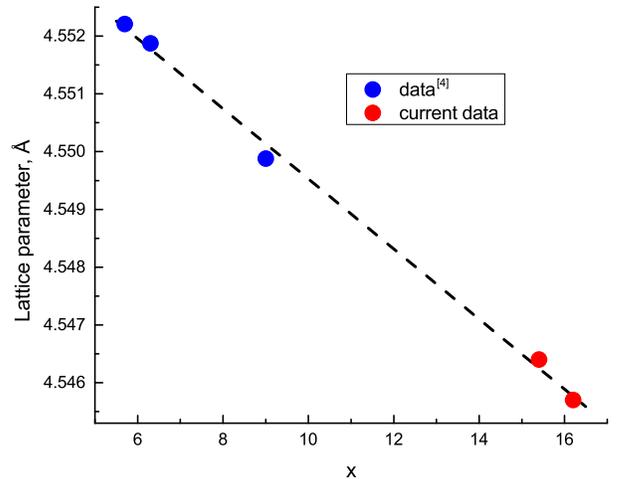}
\caption{\label{fig1} (Color online) The lattice parameters of (Mn$_{1-x}$Co$_x$)Si as a function of Co content. The observed linear dependence (Vegard rule) indicates that the Co component forms a solid solution with MnSi at the given concentrations.}
\end{figure}

Fig.~\ref{fig1} illustrates a dependence of the lattice parameters of (Mn$_{1-x}$Co$_x$)Si on the Co content. Note that the lattice parameter of the material with $x\approx 0.15-0.17$ reaches the corresponding value for pure MnSi at the quantum phase transition phase point at pressure about 14 kbar~\cite{3}.

\begin{figure}[htb]
\includegraphics[width=80mm]{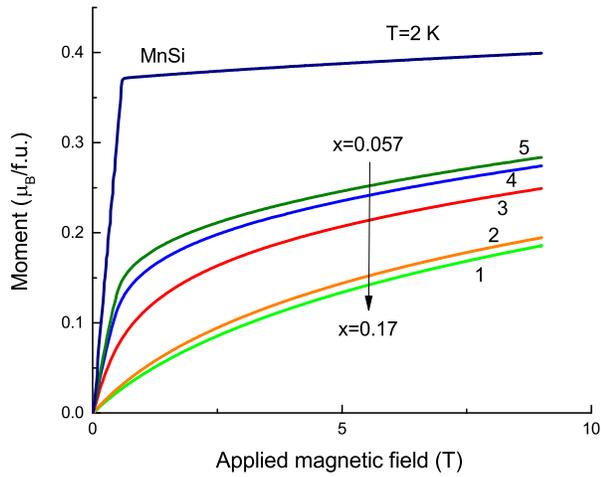}
\caption{\label{fig2} (Color online) Magnetization curves for (Mn$_{1-x}$Co$_x$)Si in comparison with one for pure MnSi. (1-5: $x=0.17, 0.15, 0.09, 0.063, 0.057$).}
 \end{figure}

The magnetization curves of (Mn$_{1-x}$Co$_x$)Si including those studied earlier~\cite{4} are shown in Fig~\ref{fig2}. The curves are naturally shifted with the Co concentration changing a form and gradually loosing saturation features. That was observed also in Ref.~\cite{2} at smaller Co concentrations.

In fact, as seen in Fig.~\ref{fig2} a saturation of magnetization does not occur in the doped samples even at magnetic fields to $9T$. It should be noted here that a spin system in heavy doped (Mn$_{1-x}$Co$_x$)Si samples at low temperatures as it will be evident further has a kind of disordered structure complicated by the intense helical fluctuations~\cite{8,9} and the Co impurity spins. Why all this prevent the magnetization from saturation remain to be seen.

\begin{figure}[htb]
\includegraphics[width=80mm]{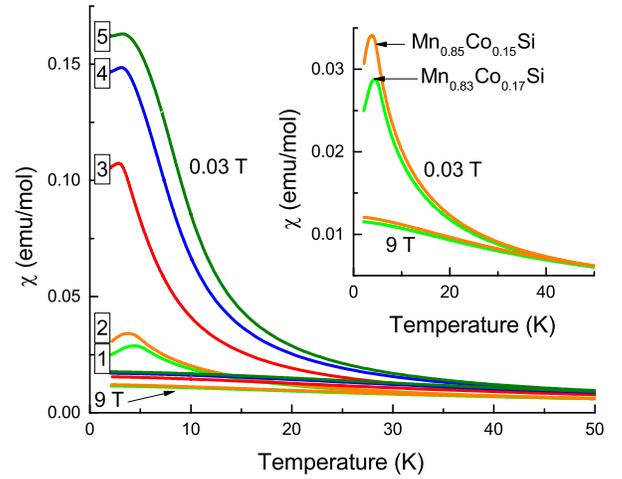}
\caption{\label{fig3} (Color online) Magnetic susceptibility of (Mn$_{1-x}$Co$_x$)Si as a function of temperature at $0.03 T$ and $9T$. See an enlarged plot of new data in the inset. (1-5: $x=0.17, 0.15, 0.09, 0.063, 0.057$).}
\end{figure}

\begin{figure}[htb]
\includegraphics[width=80mm]{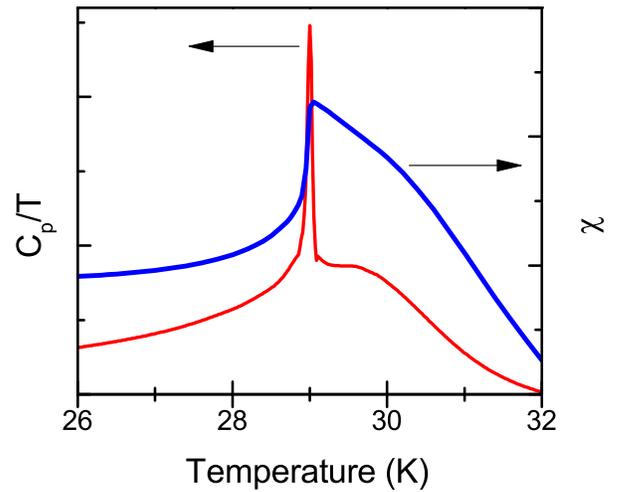}
\caption{\label{fig4} (Color online) Schematic view of magnetic susceptibility and heat capacity at the phase transition in pure MnSi. } 
\end{figure}

Nevertheless, as one can see in Fig.~\ref{fig3} absolutely clear evidences of peculiar behaviour of magnetic susceptibility that most certainly genetically connected with the inflection point on $\chi (T)$ curve of pure MnSi slightly above the phase transition point~\cite{10}. In turn the inflection point corresponds to the helical fluctuation maximum discovered at heat capacity measurements in the same range of temperature~\cite{10}(see Fig.~\ref{fig4}). So, it suggests that the helical fluctuations survive the heavily doping of MnSi with Co.

\begin{figure}[htb]
\includegraphics[width=80mm]{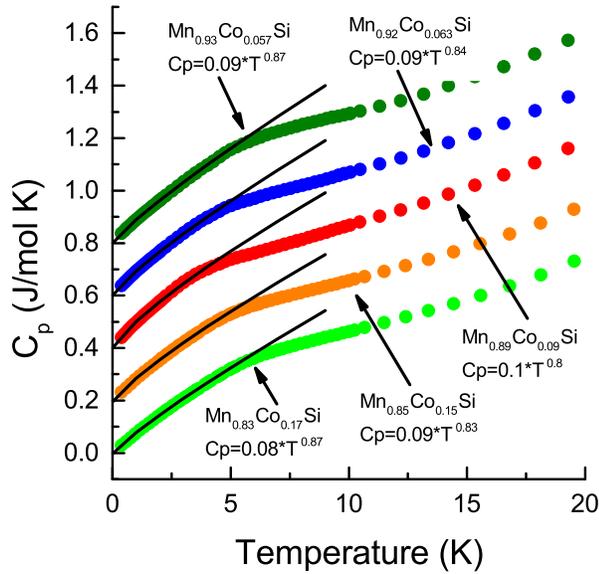}
\caption{\label{fig5} (Color online) Heat capacity of (Mn$_{1-x}$Co$_x$)Si. Fitting of the low temperature part of heat capacity to the power function is illustrated. The values of the power exponents shown in the plot. The data are shown with offsets for better viewing.}
\end{figure}

\begin{figure}[htb]
\includegraphics[width=80mm]{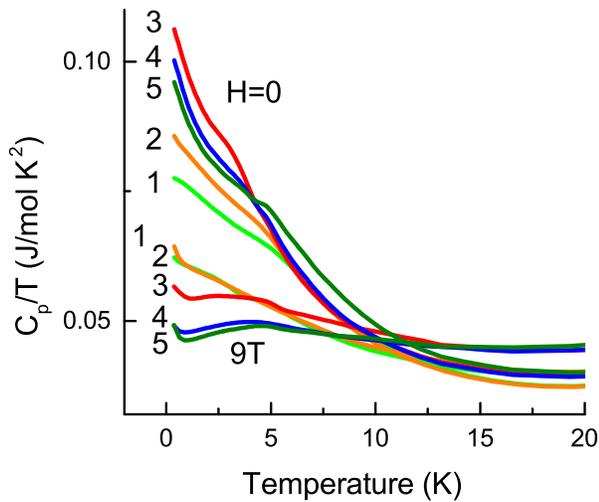}
\caption{\label{fig6} (Color online) The ratio $C_p/T$ for (Mn$_{1-x}$Co$_x$)Si samples as a function of temperature at zero and $9T$ magnetic fields. Is seen that diverging of $C_p/T$ is suppressed by strong magnetic field for samples studied earlier~\cite{4}, but this is probably not a case for two new samples with the increased concentration of Co. (1-5: $x=0.17, 0.15, 0.09, 0.063, 0.057$).} 
\end{figure}

\begin{figure}[htb]
\includegraphics[width=80mm]{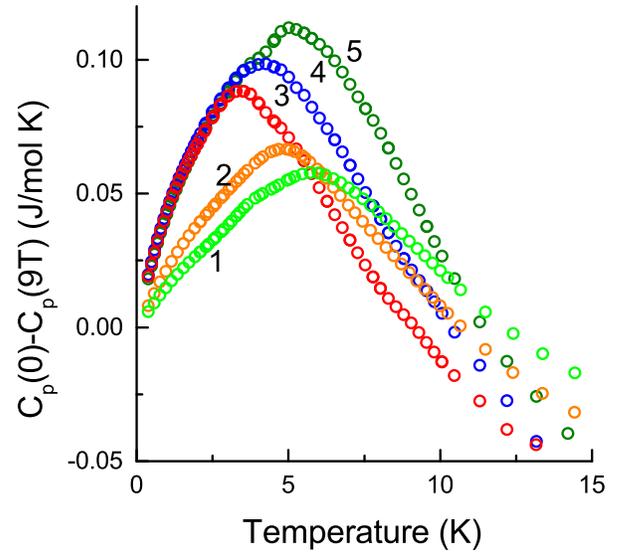}
\caption{\label{fig7} (Color online) The difference between heat capacity at zero magnetic field $C_p(0)$ and heat capacity at 9~T $C_p(9 T)$ for (Mn$_{1-x}$Co$_x$)Si samples. This manipulation implies a subtraction of some background contributions, including phonon and electron ones to the heat capacity leaving the spin fluctuation part intact. But as is seen in Fig.~\ref{fig2} magnetic field of 9~T does not suppress completely the spin fluctuations in (Mn$_{1-x}$Co$_x$)Si with x=0.15 and 0.17. That is probably why new curves look somewhat different. (1-5: $x=0.17, 0.15, 0.09, 0.063, 0.057$).}
\end{figure}

Another proofs for that follows from the heat capacity measurements shown in Fig.~\ref{fig5}, where the new results are displayed together with the data~\cite{4}.  As is seen in Fig.~\ref{fig5} the heat capacity curves noticeably change their slops at temperatures about 5~K. The low temperature parts of the curves can be described by the power function with the exponents less than one (Fig.~\ref{fig5}). This immediately leads to the diverging ratio $C_p/T$, which is a signature of quantum critical behaviour (Fig.~\ref{fig6}). The mentioned slope change becomes more evident after a subtraction from the heat capacity curve at zero magnetic field the corresponding curve at $9T$ as it was suggested in Ref.~\cite{4} (see Fig.~\ref{fig7}).

A behavior of ratio $C_p/T$ of all samples (Mn$_{1-x}$Co$_x$)Si is demonstrated in Fig.~\ref{fig6}. The diverging of $C_p/T$ at $T\rightarrow 0$ is obvious, but a weak response of samples with $x=0.15$ and $x=0.17$ to applied magnetic field may indicate an increase of the intensity of fluctuations.

\begin{figure}[htb]
\includegraphics[width=80mm]{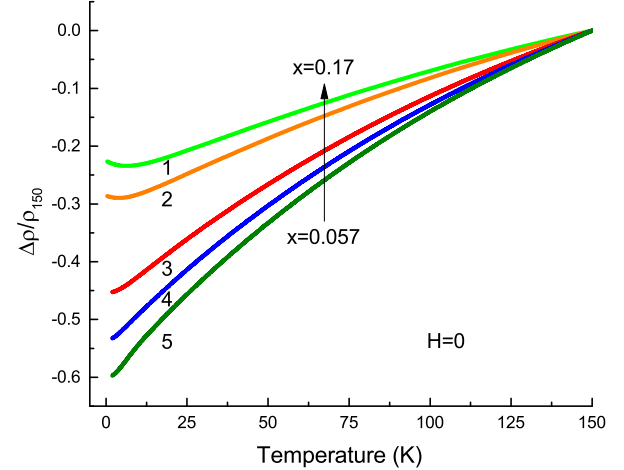}
\caption{\label{fig8} (Color online) Dependence of resistivity of (Mn$_{1-x}$Co$_x$)Si on temperature. Data are modified accordingly to the formula $\Delta R/R_{150 K}$ for better comparison. (1-5: $x=0.17, 0.15, 0.09, 0.063, 0.057$).}  
\end{figure}

\begin{figure}[htb]
\includegraphics[width=80mm]{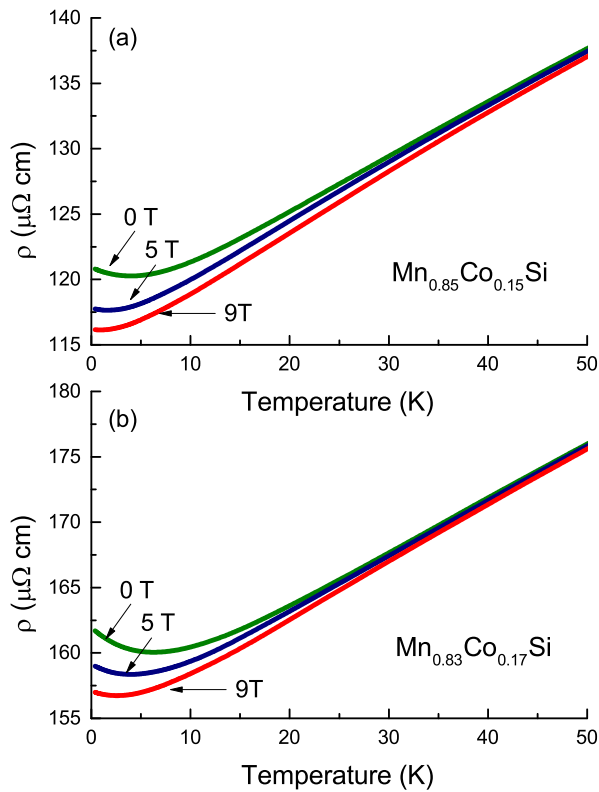}
\caption{\label{fig9} (Color online) Dependence of resistivity of (Mn$_{1-x}$Co$_x$)Si on temperature at various magnetic fields. The features of weak localization, which partly suppressed by magnetic field, are evident at low temperature.}  
\end{figure}

\begin{figure}[htb]
\includegraphics[width=80mm]{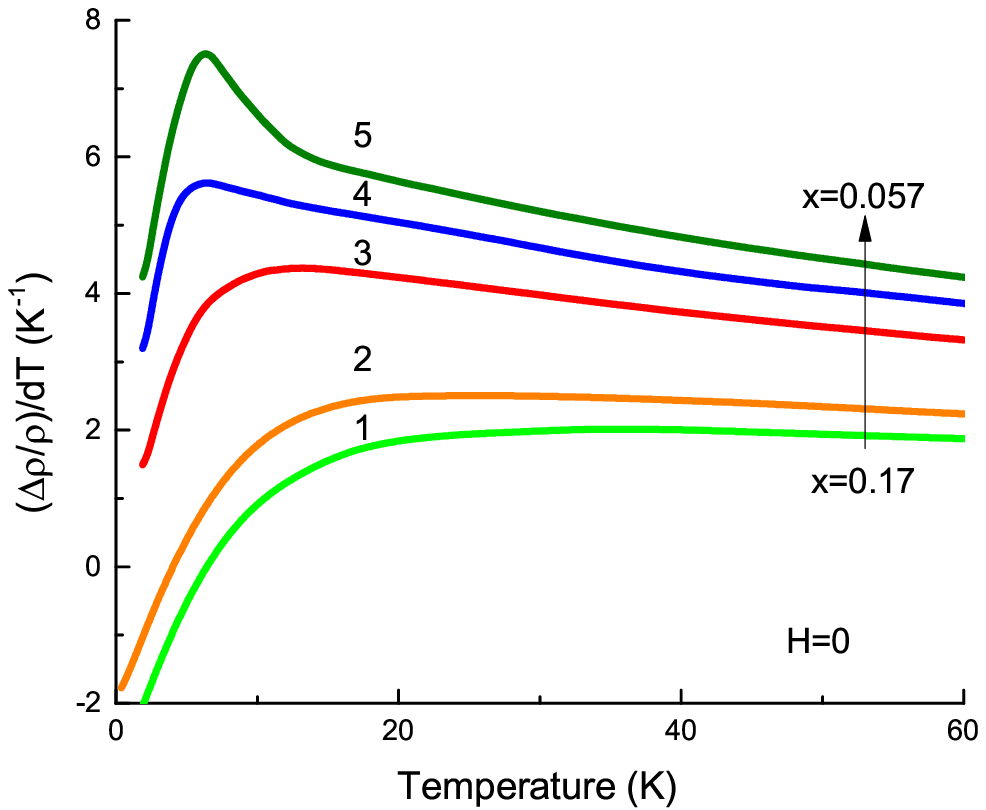}
\caption{\label{fig10} (Color online) Temperature derivatives of resistivity of the samples (Mn$_{1-x}$Co$_x$)Si. Note negative values of the derivatives for $x=0.15$ and $x=0.17$ at low temperatures. (1-5: $x=0.17, 0.15, 0.09, 0.063, 0.057$). } 
\end{figure}

The resistivity measurements are illustrated in Fig.~\ref{fig8}--\ref{fig10}.

How as seen in Fig.~\ref{fig8} the temperature dependence of resistivity of samples becomes reduced with the doping increase that is actually expected. But the weak localization features for samples with $x=0.15$ and $x=0.17$ (see Fig.~\ref{fig9}) were not anticipated on the basis of heat capacity data, which show no distinct differences between the samples. So, the conductivity electrons probably do not contribute to the magnetic fluctuation phenomena.  

Temperature derivatives of resistivity of the samples (Mn$_{1-x}$Co$_x$)Si depicted in Fig.~\ref{fig10} show progressive degradation of the specific features of corresponding curves with doping therefore probably indicating that the electron scattering on magnetic fluctuations becomes dominated by the impurity scattering.

\begin{figure}[htb]
\includegraphics[width=80mm]{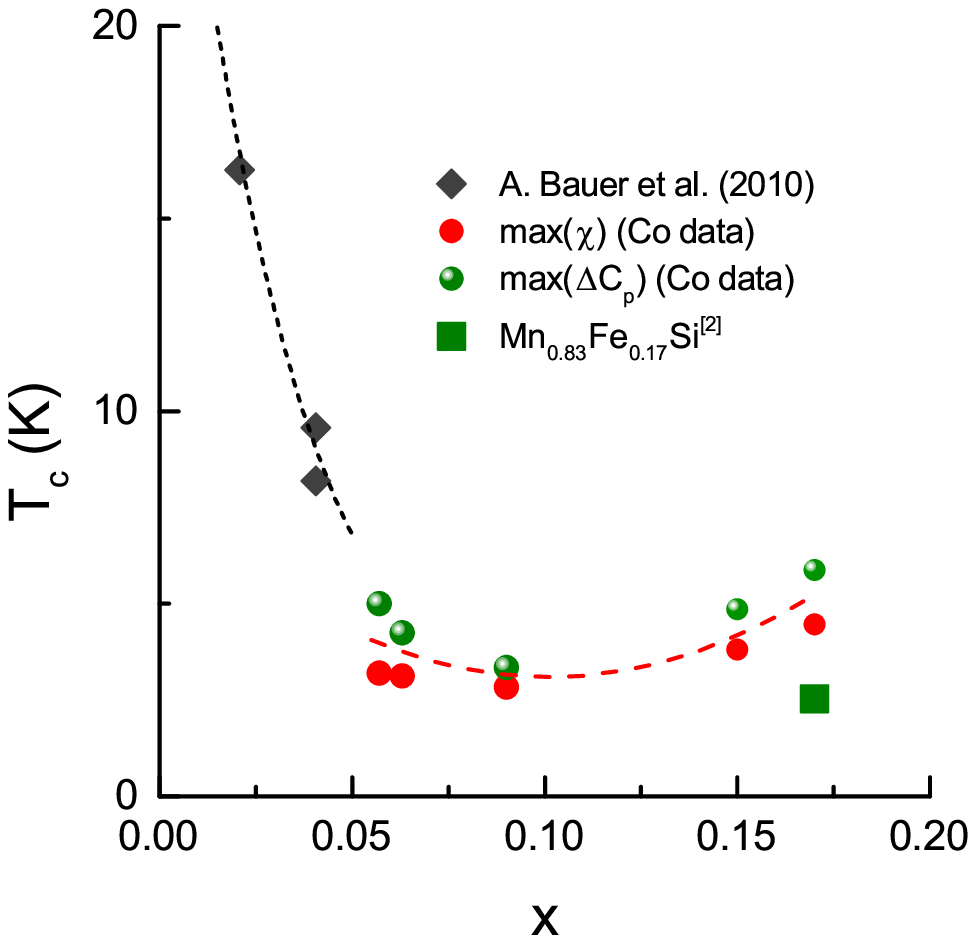}
\caption{\label{fig11} (Color online) Temperature of fluctuation maxima as a function of Co concentration. Phase transition temperature occurring at low Co content is also shown as a black squares~\cite{2}. In Ref.~\cite{2} phase transition temperatures were determined from the temperature dependence of magnetic moments. Current fluctuation maxima data were taken as the $\chi$ maxima (Fig.~\ref{fig3}) and maxima of $\Delta C_p$ (Fig.~\ref{fig7}). Note that the iron square data point corresponds to the sample studied in Ref.~\cite{10}. Corrected composition was used. Coordinates of the data points with $x=0.15$ and $x=0.17$ calculated from the heat capacity obviously influenced by the problem pointed out in the caption of Fig.~\ref{fig7}.} 
\end{figure}

\begin{figure}[htb]
\includegraphics[width=80mm]{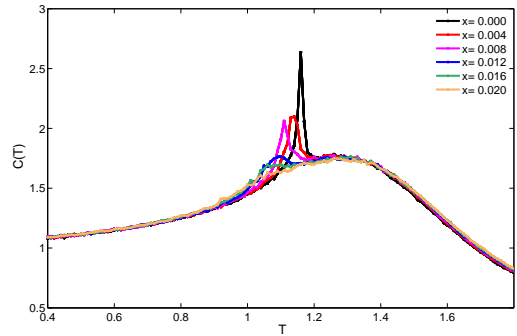}
\caption{\label{fig12} (Color online) Monte-Carlo modelling the heat capacity behavior of a classical spin system with impurities situated in interstitial positions for different doping concentrations~\cite{11}.} 
\end{figure}

We can see in Fig.~\ref{fig11} that the helical fluctuations do not disappear on heavy doping with Co. Moreover, the temperature of fluctuation maxima practically does not change with Co concentration as a contrast to a behavior of the phase transition temperature. So, we have to confirm our former conclusion that that at large concentration of Co substitutes a cloud of the helical fluctuations spreading over a significant range of concentrations and temperatures arose close to 0~K (Ref.~\cite{4}). Some insight can be obtained from Fig.~\ref{fig12}, where a result of the Monte-Carlo modelling of behaviour of a classical spin system with the frozen impurities in interstitial positions is shown \cite{11}. As is seen the helical phase transition in this spin system disappears with doping whereas the fluctuation maximum is not changed at all. Probably just a frozen nature of impurities not an exact their positions are responsible for such behavior of fluctuations.

It needs to be reminded that as a difference to the spin model our real (Mn$_{1-x}$Co$_x$)Si system experience a regular volume decrease on doping (see Fig.~\ref{fig1}) that is equivalent to hydrostatic compression of the material. So, if we would try to describe our finding (Fig.~\ref{fig11}) in terms of temperature and pressure it would look like the diagram depicted in Fig.~\ref{fig13}. The suggested extended region of helical fluctuations in Fig.~\ref{fig13} probably correlates with early findings of the non-Fermi liquid behaviour~\cite{12} and the partial order~\cite{13} in pure MnSi at pressure above the magnetic phase transition and low temperature.

\section{Summary and conclusion}
1. The observed linear dependence of the lattice parameters of (Mn$_{1-x}$Co$_x$)Si (Vegard rule) indicates that the Co component forms a solid solution with MnSi at the studied concentrations (Fig.~\ref{fig1}).

2. A saturation of magnetization does not occur in the doped samples to $9T$ (Fig.~\ref{fig2}). This correlates with a weak response of the diverging part of $C_p/T$ to strong magnetic field.

3.	The helical fluctuations survive the heavily doping of MnSi with Co (Fig.~\ref{fig3}).

4. The low temperature parts of the heat capacity curves can be described by the power function with the exponents less than one. This immediately leads to the diverging ratio $C_p/T$, which is a signature of quantum critical behaviour (Fig.~\ref{fig5}).

5. The diverging of $C_p/T$ at $T\approx 0$ is clearly observed, but a weak response of samples with $x=0.15$ and $x=0.17$ to applied strong magnetic field looks surprising though perhaps indicates a change in the intensity of helical interaction (Fig.~\ref{fig6}).

6. The weak localization features for samples with $x=0.15$ and $x=0.17$ (see a Fig.~\ref{fig9}) were observed.

7. Temperature derivatives of resistivity of the samples (Mn$_{1-x}$Co$_x$)Si show progressive degradation of the specific features of corresponding curves with doping therefore probably indicating that the electron scattering on magnetic fluctuations becomes dominated by the impurity scattering (Fig.~\ref{fig10}).

8. Helical fluctuations do not disappear on heavy doping with Co. Moreover, the temperature of fluctuation maxima practically does not change with Co concentration as a contrast to a behaviour of the phase transition temperature (Fig.~\ref{fig11}).

\begin{figure}[htb]
\includegraphics[width=80mm]{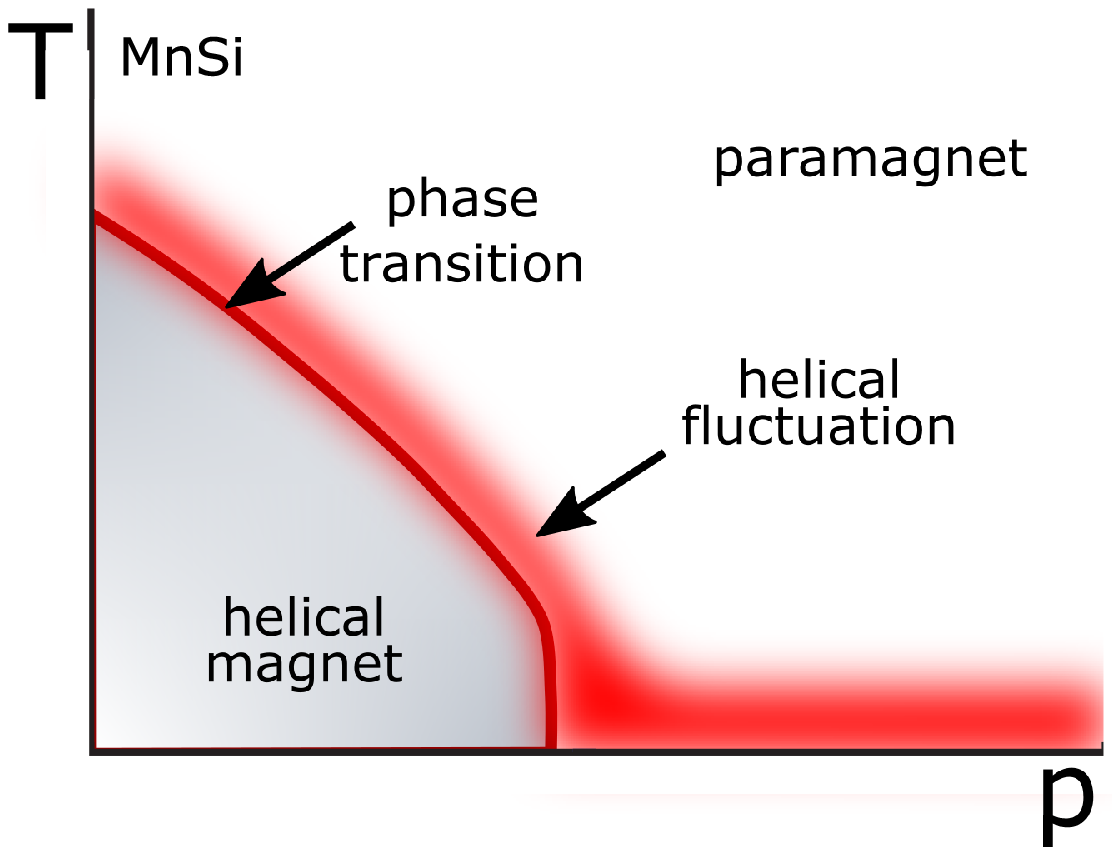}
\caption{\label{fig13} (Color online) Tentative T-P phase diagram for an itinerant magnet with the Dzyaloshinskii-Moriya interaction. } 
\end{figure}
  
In conclusion we suggest a P--T phase diagram of an itinerant magnet with an extended helical fluctuation area, which may provide certain insight to the non Fermi liquid behaviour and partial order of MnSi at high pressure and low temperature~\cite{12,13} (Fig.~\ref{fig13}).

\section{Acknowledgements}
The authors gratefully acknowledge the technical support of Dr. V.A. Sidorov.

\end{document}